\documentclass[preprint,aps,floatfix,superscriptaddress,pre,showpacs]{revtex4}
\usepackage{amsmath, amsthm, amssymb, graphicx}
\input epsf.sty
\begin{document}

\def\g{\gamma}
\def\r{\rho}
\def\w{\omega}
\def\wo{\w_0}
\def\wp{\w_+}
\def\wm{\w_-}
\def\t{\tau}
\def\av#1{\langle#1\rangle}
\def\pf{P_{\rm F}}
\def\pr{P_{\rm R}}
\def\F#1{{\cal F}\left[#1\right]}

\title{Surface critical exponents at a discontinuous bulk transition}

\author{Linjun Li and Michel Pleimling}
\affiliation{Department of Physics, Virginia Tech, Blacksburg, Virginia 24061-0435, USA}

\date{\today}

\begin{abstract}
Systems with a bulk first-order transition can display diverging correlation lengths close
to a surface. This surface induced disordering yields a special type of surface criticality.
Using extensive numerical simulations we study surface quantities in the two-dimensional 
Potts model with a large number of states $q$ which undergoes a discontinuous bulk transition. 
The surface critical exponents are thereby found to 
depend on the value of $q$, which is in contrast to prior claims that these exponents
should be universal and independent of $q$. It follows that surface induced disordering at first-order transitions
is characterized by exponents that depend on the details of the model.
\end{abstract}

\maketitle

\section{Introduction}
The presence of a surface is well known to change locally the critical properties at an
equilibrium critical point \cite{Bin83,Die86,Dos92,Die97,Ple04}. At the so-called ordinary transition, where only the bulk
correlation length diverges, this breaking of spatial translation invariance yields 
a surface universality class characterized by a set of critical exponents that 
differ from the corresponding bulk critical exponents. This change of local critical
exponents due to the presence of a surface is not restricted to equilibrium phase transitions,
but has recently also been revealed at a dynamic phase transition \cite{Par12}. In addition,
enhanced surface couplings in three-dimensional systems may result in a variety of surface phase 
transitions.
Whereas at the special
point both the bulk and the surface are critical, at the so-called surface transition only the surface
orders whereas the bulk remains disordered. Finally, at the extraordinary transition the 
bulk orders in presence of an ordered surface. Additional phenomena, as for example critical wetting,
are encountered when external fields are applied \cite{Die88}.

A special kind of surface critical phenomenon can be encountered in situations where the bulk undergoes
a first-order phase transition \cite{Lip82,Lip83,Lip83a,Lip84,Lip87}. 
Due to surface induced disordering the surface order parameter may go
continuously to zero when one approaches the bulk transition temperature from below, i.e. one has a first-order
transition in the bulk but a second-order transition at the surface. This surface critical behavior,
which is coupled to an interface delocalization phenomenon, is characterized by a set of surface
critical exponents. This general scenario has been verified for a range of systems as for example an
effective interface model \cite{Lip83b}, face-centered-cubic Ising antiferromagnets \cite{Gom88,Sch96},
body-centered-cubic alloys \cite{Hel90,Haa00} or two-dimensional Potts models undergoing a discontinuous
bulk phase transition \cite{Lip82a,Igl99}. 

Related surface (or, more appropriately, interface) phenomena are also encountered in non-equilibrium situations 
where magnetic systems are sheared or magnetic blocks are moved passed each other, 
thereby giving rise to magnetic friction \cite{Kad08,Huc09,Igl11,
Ang12,Huc12,Li13}. When the involved bulk systems
have an equilibrium first-order transition, as it is the case for two-dimensional and three-dimensional Potts system
with a large number of states, non-equilibrium surface transitions of different types (continuous, discontinuous, tricritical)
emerge \cite{Igl11,Li13}.

Early on, the values of the different local exponents at surface induced disordering
were determined within Landau theory \cite{Lip82,Lip83,Lip83a}.
Surface exponents were also determined in an effective interface model that incorporates
fluctuations by expanding around the mean-field solution \cite{Lip83b}.
However, until now only very few attempts have been made to
compute for specific systems the surface critical exponents at a discontinuous bulk transition.
For instance, in numerical simulations of both face-centered-cubic \cite{Sch96}
and body-centered-cubic \cite{Haa00} alloys nonuniversal exponents were observed at surface induced
disordering. In a study \cite{Igl99} of the two-dimensional Potts model, which undergoes a first-order
phase transition when the number of states $q$ is larger than four, the surface exponents were computed
exactly in the $q \longrightarrow \infty$ limit. In addition, systems with $q =7$, 8, or 9 states were studied numerically
for small stripes with mixed boundary conditions using DMRG. Extrapolating to infinity the data obtained for stripes with
width less than 40 rows, it was concluded that the corresponding surface exponents
were universal and identical to those of the $q \longrightarrow \infty$ case.

In this paper we are revisiting surface induced disordering in the two-dimensional Potts model. Using extensive numerical
simulations, we study large systems with up to $q=100$ states. Our numerical data for a variety of quantities
and boundary conditions show in a consistent way that the surface exponents at the bulk first-order transition
depend on the number of states. Therefore, while the general scenario of surface
induced disordering is correct, the phenomenon does not yield universal surface exponents, but instead the values
of these exponents are nonuniversal and depend on microscopic details like the number of states.

Our paper is organized in the following way. We introduce the model and the quantities of interest 
in the next Section, before briefly reviewing in Section III
some of the general results obtained for surface induced disordering, as far as they are relevant
for our study. In Section IV we present our numerical results and obtain the values of different 
surface exponents. As we show, the different quantities yield a consistent picture
indicating that the surface exponents at the first-order transition in the two-dimensional
Potts model depend on the number of states and are therefore not universal.
Finally, in Section IV we discuss our results and conclude.

\section{Model and quantities}

In the following we study the ordering of the two-dimensional Potts model in presence of a surface, characterized
by the Hamiltonian
\begin{equation}
{\mathcal H} = - J_b \sum\limits_{bulk} \delta(s_{x,y}- s_{x',y'}) - J_s \sum\limits_{surface} 
\delta(s_{x,y_s}-s_{x',y_s})~,
\end{equation}
where $\delta(\cdots)$ is the Kronecker delta, with $\delta(x)=1$ if $x=0$ and $\delta(x)=0$ otherwise. In this Hamiltonian
the first term describes the interaction between nearest neighbor Potts spins where at least one of them does not belong
to the surface, whereas the second term is the interaction between nearest neighbor spins that are both located at the surface
row $y_s$. The ferromagnetic
bulk and surface coupling constants are given by $J_b > 0$ and $J_s > 0$. Throughout this study we only
consider the case $J_s = J_b = J$. Temperatures are measured in units such that $J/k_B =1$ where $k_B$ is the
Boltzmann constant.

Our focus is on cases where the bulk transition is of first order \cite{Wu82}. For that reason we consider at every
lattice site $(x,y)$ a Potts spin $s_{x,y}$ that can take on $q$ values, $s_{x,y} = 1, \cdots, q$, with $q > 4$.
We thereby probe a wide range of $q$ values, namely $q=5$, 9, 16, and 100.

We consider systems composed of $L \times M$ spins, where the values of $L$ and $M$ used in our study will be discussed
below. We always consider periodic boundary conditions in the $x$-direction. Most of our quantities are obtained
for free boundaries in the $y$-direction, yielding systems with two surfaces located at $y=1$ and $y=M$. In order
to make a connection with the data discussed in \cite{Igl99}, we also investigate some systems with symmetry-breaking
boundary conditions where the spins along one of the surfaces are at a fixed common value.

Due to the presence of surfaces, local quantities depend on the row index $y$
\cite{Bin83,Die86,Dos92,Die97,Ple04}.
Our quantities of interest include the local magnetization, $m(y)$, the surface excess magnetization, $m_s$,
as well as the spin-spin correlations along and perpendicular to the surface, $C_{ss}(x)$ and $C_{sb}(y)$, respectively,
where the index $s$ ($b$) indicates a surface (bulk) spin.

The magnetization of row $y$ is given by
\begin{equation}
m(y) = (q N_m(y) /L -1)/(q-1)
\end{equation}
where $N_m(y)$ is the average number of majority spins in that row: $N_m(y) = \mbox{max}(N_1(y), \cdots, N_q(y))$, 
with $N_k(y)$ being the
average number of spins in state $k$ in row $y$. The surface magnetization $m_1$ is given by $m_1 = m(1)$ (which equals $m(M)$ in
the case of free boundary conditions). From the magnetization profile we can derive the so-called surface excess magnetization
associated with the surface located at $y=1$:
\begin{equation} \label{ms}
m_s = \sum\limits_{y=1} \left( m_b - m(y) \right)~,
\end{equation}
where $m_b$ is the bulk magnetization. When approaching the $q$-dependent bulk transition temperature 
$T_c(q) = 1/\ln ( 1 + \sqrt{q})$ from below,
both the surface magnetization and the surface excess magnetization are displaying close to $T_c(q)$ an algebraic behavior
as a function of temperature \cite{Lip82,Lip83a}:
\begin{eqnarray}
m_1 & \sim & \left[ (T_c(q) - T)/T_c(q) \right]^{\beta_1} \\
m_s & \sim & \left[ (T_c(q) - T)/T_c(q) \right]^{\beta_s}
\end{eqnarray}
with the critical exponents $\beta_1$ and $\beta_s$ respectively.

Surface induced disordering is an anisotropic critical phenomenon, governed by two different correlation
length exponents $\nu_\parallel$ and $\nu_\perp$ for correlations parallel and perpendicular to
the surface \cite{Lip83}. The corresponding temperature-dependent
correlation lengths $\xi_\parallel$ and $\xi_\perp$ can be
obtained from the surface-surface and surface-bulk correlators:
\begin{eqnarray}
C_{ss}(x) = \frac{q}{q-1} \left< \delta \left(s_{1,1}- s_{x,1} \right)- \frac{1}{q} \right> - m_1^2 \sim \exp(- x /\xi_\parallel) \label{Css} \\
C_{sb}(y) = \frac{q}{q-1} \left< \delta \left(s_{1,1}- s_{1,y} \right)- \frac{1}{q} \right> - m_1 m(y) \sim \exp(- y/\xi_\perp) \label{Csb}
\end{eqnarray}
with
\begin{eqnarray}
\xi_\parallel & \sim & \left[ (T_c(q) - T)/T_c(q) \right]^{- \nu_\parallel} \label{xipar} \\
\xi_\perp & \sim & \left[ (T_c(q) - T)/T_c(q) \right]^{- \nu_\perp} \label{xiperp}
\end{eqnarray} 
close to $T_c(q)$.

Our aim in the following is to obtain values for the critical exponents $\beta_1$, $\beta_s$,
$\nu_\parallel$, and $\nu_\perp$ as a function of the number of states $q$. In order to do so
we combine two complementary approaches. In the first approach we determine our quantities as a
function of temperature and calculate from these data effective exponents that yield the critical
exponents in the limit $T \longrightarrow T_c(q)$ \cite{Ple04,Ple98}. For example, for the surface
magnetization $m_1$ we obtain the effective surface exponent via the logarithmic derivative
\begin{equation} \label{beff}
\beta_{1,eff}(t) = \frac{d \ln m_1}{d \ln t}
\end{equation}
with the reduced temperature $t = (T_c(q) - T)/T_c(q)$. We then have
\begin{equation}
\beta_1 = \lim\limits_{t \longrightarrow 0} \beta_{1,eff}(t)~.
\end{equation}
For this approach only data not affected by finite-size effects should be used, which requires the
comparison of systems of different sizes. For that reason we study large ranges of $L$ and $M$, 
ranging from $L = 80$ to $L=1280$ and from $M = 40$ to $M=320$. In the second approach we try to use 
finite-size effects to our advantage. Following \cite{Igl99}, we consider the surface magnetization
at the exactly known bulk transition temperature $T_c(q)$, using symmetry-breaking boundary conditions.
For a system with $L = \infty$ and width $M$, the surface magnetization should then depend algebraically
on the scaling dimension $x_1 = \beta_1/\nu_\perp$, i.e. $m_1 \sim M^{-x_1}$. In order to mimic a stripe of infinite length we
study systems with $L$ up to $L =64000$. As a drawback, only stripes with rather small widths $M$ 
can be simulated. Still, comparing our estimates
for the scaling dimension with the independently determined values of $\beta_1$ and $\nu_\perp$ using 
temperature-dependent quantities provides us with an important check whether our different data sets are
consistent.

\section{Scaling in presence of surface induced disordering}

When the bulk undergoes a continuous phase transition, as it is for example the case for the Ising model,
then the surface critical exponents and the relationship between these exponents can be derived from
the scaling form of the singular part of the surface free energy \cite{Bin83,Die86,Dos92}.
As shown in \cite{Lip83} a similar phenomenology holds in the case of surface induced disordering.
Indeed, as function of the two scaling fields $t$ (the reduced temperature) and $h_1$ (a surface field),
the singular part of the surface free energy may be written as \cite{Lip83}
\begin{equation}
f_s = | t|^{2 - \alpha_s} \Omega(|t|^{-\Delta_1} h_1)
\end{equation}
with the two independent surface exponents $\alpha_s$ and $\Delta_1$. All surface critical exponents 
can then be expressed through these two independent exponents. Specifically for some of the exponents at the
center of our study one obtains \cite{Lip83}:
\begin{eqnarray}
\beta_1 & = & 2 - \alpha_s - \Delta_1 \label{exp1} \\
\beta_s & = & 1 - \alpha_s \label{exp2} \\
(d-1) \nu_\parallel & = & 2 - \alpha_s \label{exp3}
\end{eqnarray}
Combining Equations (\ref{exp2}) and (\ref{exp3}) yields for two-dimensional systems with $d=2$ the relationship
\begin{equation} \label{consistency}
\nu_\parallel = 1 + \beta_s
\end{equation}
which provides an additional check for the consistency of our data.

\section{Results}

%%%%%%%%%%%%%%%%%%%%%%%%%%%%%%%%%%%%%%%%%%%FIG 1.%%%%%%%%%%%%%%%%%%%%%%%%%%%%%%%%%%%%%%%%%%%%%%%%%%%%%%
\begin{figure} [h]
\includegraphics[width=0.80\columnwidth]{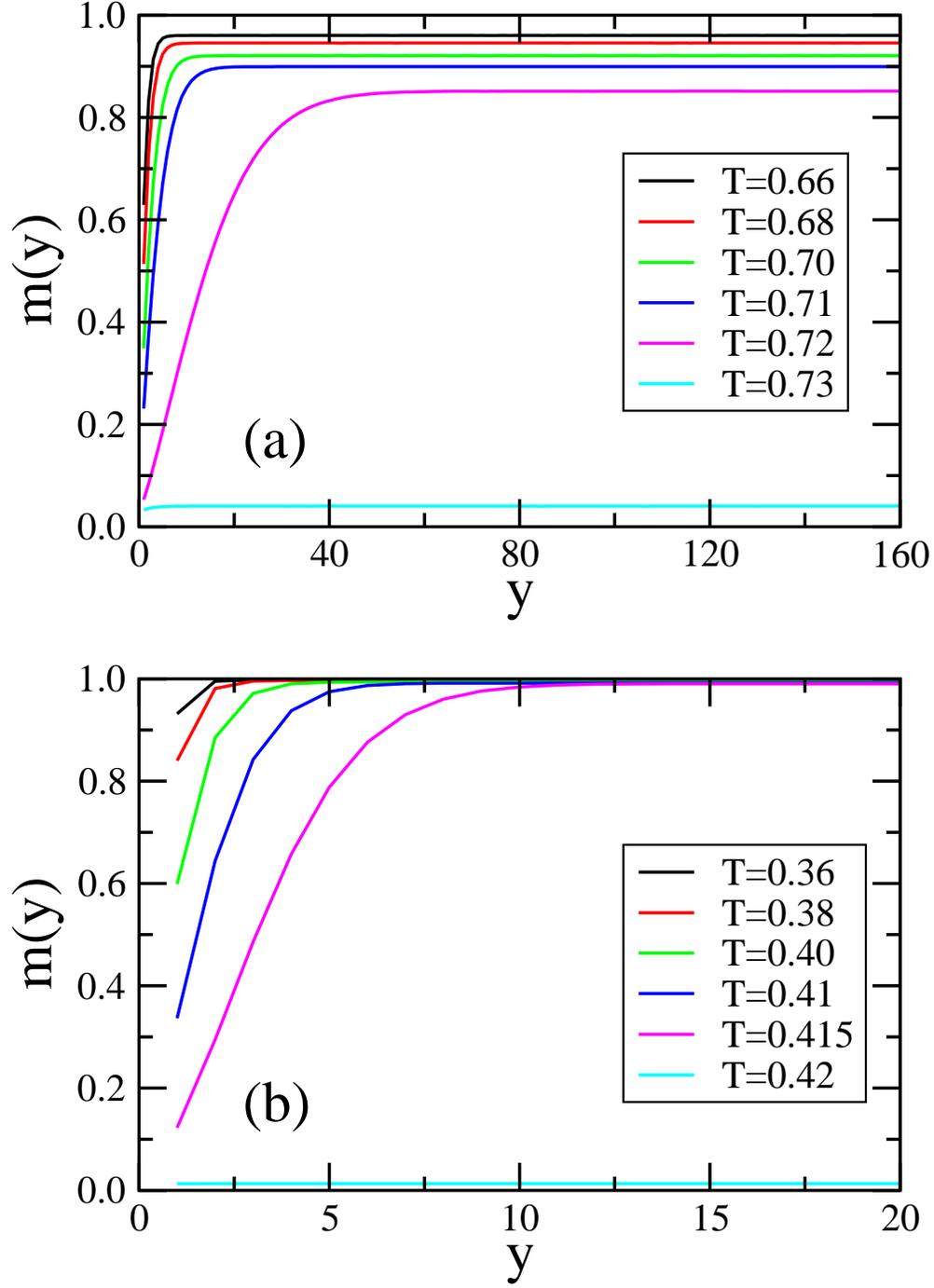}
\caption{\label{fig1} (Color online)
Layer magnetization density as a function of layer index $y$ for (a) $q=9$ and (b) $q=100$ states. The layer
$y=1$ is the surface layer. Approaching the bulk transition temperature from below, the surface magnetization
decreases continuously whereas deep inside the bulk the local magnetization displays a discontinuous jump.
The highest temperature included is just above the corresponding bulk transition temperature.
The data have been obtained for a system composed of $640 \times 320$ spins. Error bars, that result from averaging over
typically ten independent runs, are of the order of the
thickness of the lines. 
}
\end{figure}
%%%%%%%%%%%%%%%%%%%%%%%%%%%%%%%%%%%%%%%%%%%FIG 1.%%%%%%%%%%%%%%%%%%%%%%%%%%%%%%%%%%%%%%%%%%%%%%%%%%%%%%

Due to the surface induced disordering, bulk and surfaces behave remarkably different when approaching
the first-order bulk transition temperature: whereas the bulk magnetization displays a discontinuous jump
at that temperature, the surface magnetization vanishes continuously. This is illustrated in Fig. \ref{fig1} 
for two different number of states $q$ through
the magnetization profiles close to a surface of systems composed of $640 \times 320$ spins.

%%%%%%%%%%%%%%%%%%%%%%%%%%%%%%%%%%%%%%%%%%%Tab. 1%%%%%%%%%%%%%%%%%%%%%%%%%%%%%%%%%%%%%%%%%%%%%%%%%%%%%%
\begin{table}[h] \label{table1}
\caption{Surface critical exponents for two-dimensional Potts models with $q$ states that have a first-order bulk
transition.}
\begin{tabular}{|c||c|c|c|c|}
\hline
$q$ & $\beta_1$ & $\beta_s$ & $\nu_\parallel$ & $\nu_\perp$ \\
\hline
5 & $0.64(2)$ & $-0.63(3)$ & $0.42(3)$ & $0.56(5)$ \\
9 & $0.79(1)$ & $-0.54(2)$ & $0.53(4)$ & $0.46(2)$ \\
16 & $0.90(2)$ & $-0.46(3)$ & $0.55(3)$ & $0.43(2)$ \\
100 & $0.95(2)$ & $-0.40(2)$ & $0.59(2)$ & $0.38(2)$ \\
\hline
$\infty$ \cite{Igl99}  & $1$ & $-$ & $2/3$ & $1/3$ \\
\hline
\end{tabular}
\end{table}
%%%%%%%%%%%%%%%%%%%%%%%%%%%%%%%%%%%%%%%%%%%Tab. 1%%%%%%%%%%%%%%%%%%%%%%%%%%%%%%%%%%%%%%%%%%%%%%%%%%%%%%

\begin{figure} [h]
\includegraphics[width=0.80\columnwidth]{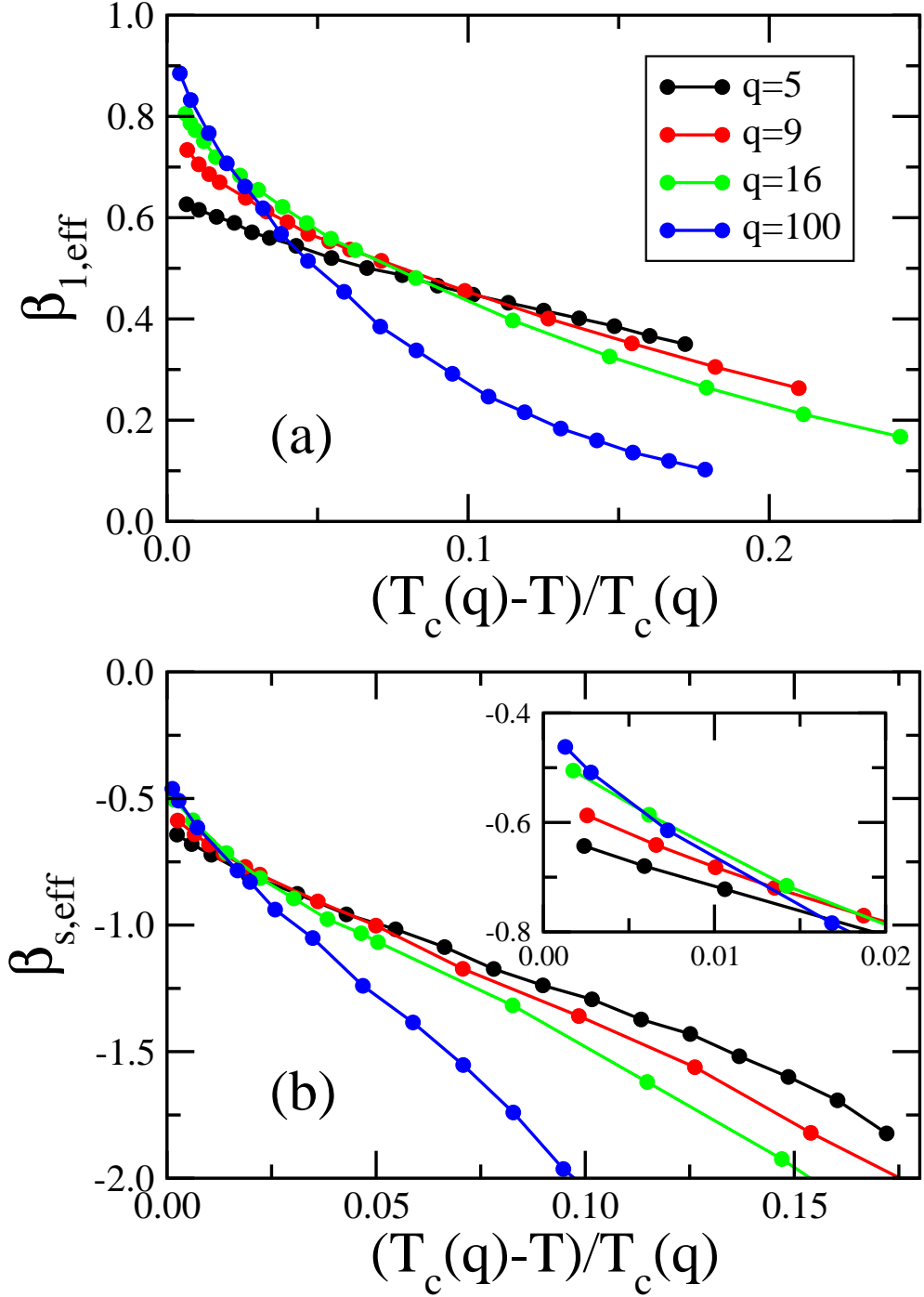}
\caption{\label{fig2} (Color online)
(a) Effective exponent of the surface magnetization and (b) effective exponent of the surface excess magnetization 
as function of the reduced temperature $(T_c(q)-T)/T_c(q)$ for systems with different numbers of states $q$. Here,
$T_c(q)$ is the bulk transition temperature for the model with $q$ states. The inset in (b) shows a blow-up
close to the transition temperature. The values of the exponents obtained from extrapolating to the transition
temperature are given in Table I. Only data not affected by finite-size effects are shown. Error bars
are of the order of the symbol sizes in the main figures.
}
\end{figure}

Data like that shown in Fig. 1 allow us to obtain the temperature dependence of both the surface magnetization
and the surface excess magnetization (\ref{ms}). From the logarithmic derivative with respect
to the reduced temperature $t = (T_c(q) - T)/T_c(q)$, see Eq. (\ref{beff}), we obtain effective exponents that in the 
limit $t \longrightarrow 0$ yield the values of the critical exponents. The result of this procedure is shown
in Fig. \ref{fig2} for (a) the surface magnetization and (b) the surface excess magnetization. As explained in Section II,
the use of temperature-dependent effective exponents is only meaningful when data for the semi-infinite system
are studied. For that reason we very carefully analyzed finite-size effects by comparing data for a range of different
system sizes. Only data that were identical for at least two different system sizes have been retained.
The curves shown in Fig. \ref{fig2} for the different values of $q$ are therefore free of any finite-size effects.
We first note that the effective exponents do not display a plateau but instead keep
changing as a function of the reduced temperature. This is similar to what one observes in situations where the bulk undergoes
a second-order phase transition \cite{Ple98}. Extrapolating the data to $t=0$, we obtain the values
for the critical exponents listed in Table I. We note that the exponent $\beta_s$ of the surface excess magnetization
is negative, in agreement with a diverging disordered region at the bulk transition temperature.

%%%%%%%%%%%%%%%%%%%%%%%%%%%%%%%%%%%%%%%%%%%FIG 2.%%%%%%%%%%%%%%%%%%%%%%%%%%%%%%%%%%%%%%%%%%%%%%%%%%%%%%
%%%%%%%%%%%%%%%%%%%%%%%%%%%%%%%%%%%%%%%%%%%FIG 2.%%%%%%%%%%%%%%%%%%%%%%%%%%%%%%%%%%%%%%%%%%%%%%%%%%%%%%

As our systems are rather large, we manage to obtain data not affected by finite-size effects up to very
close to the phase transition point. Extrapolating the data to the phase transition point itself using different
functions yield slightly different values for the critical exponents. Our error bars in Table I are conservative
and take these differences in the extrapolated values into account.

The results shown in Fig. \ref{fig2} and Table I indicate that the surface exponents depend on the
number of states $q$ and that their values approach the exactly known value for $q \longrightarrow \infty$ \cite{Igl99}
when $q$ increases. This scenario is different to that discussed in \cite{Igl99} where it was claimed that
the surface critical exponents are independent of $q$ and identical to the $q \longrightarrow \infty$ case.
As we discuss in the following, our data show in a consistent way that the Potts surface critical exponents at 
a first-order bulk transition are indeed not universal.

%%%%%%%%%%%%%%%%%%%%%%%%%%%%%%%%%%%%%%%%%%%FIG 3.%%%%%%%%%%%%%%%%%%%%%%%%%%%%%%%%%%%%%%%%%%%%%%%%%%%%%%
\begin{figure} [h]
\includegraphics[width=0.80\columnwidth]{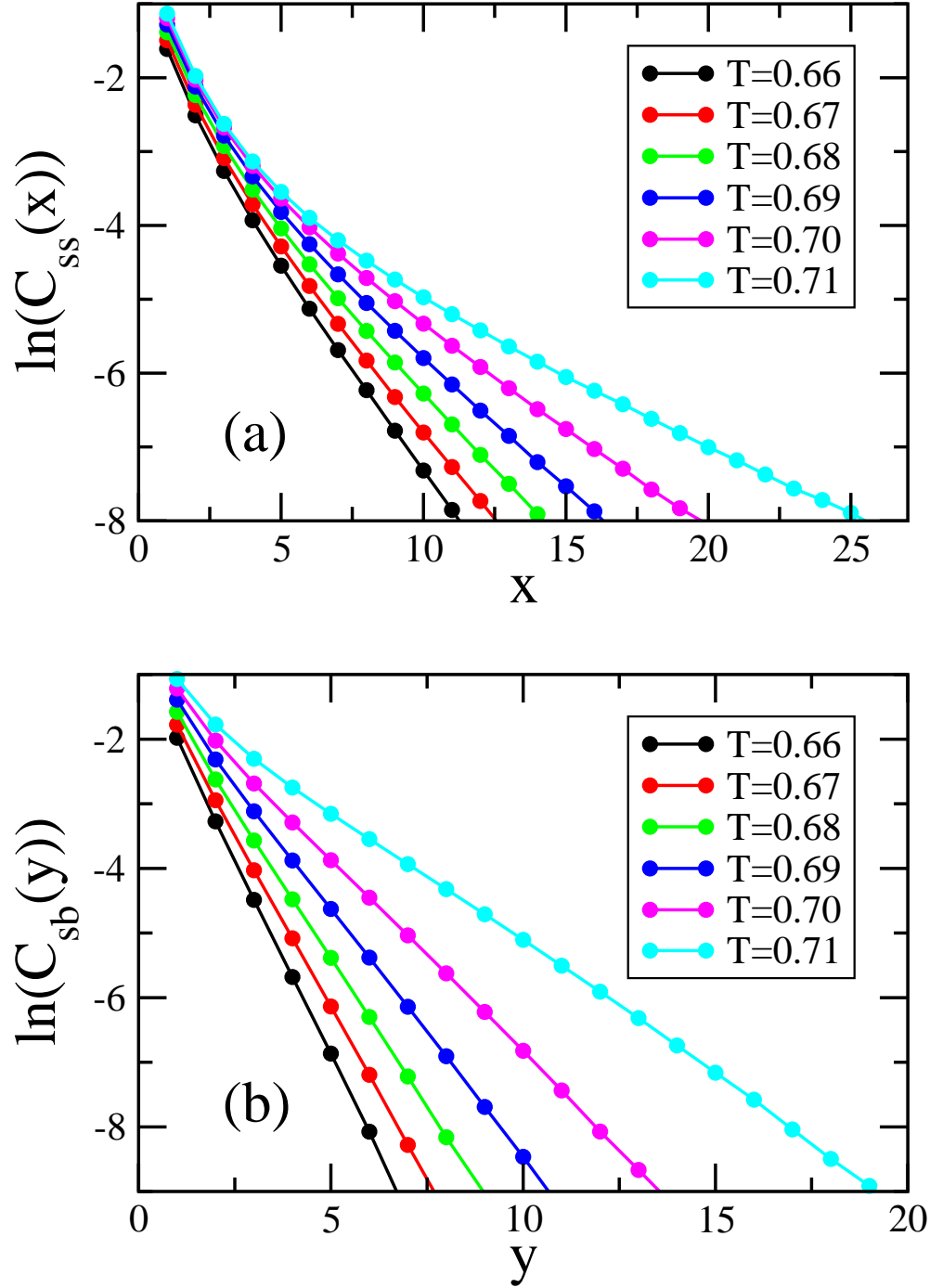}
\caption{\label{fig3} (Color online)
(a) Surface-surface and (b) surface-bulk correlation for the $q=9$ case and different temperatures below
the bulk transition temperature. Both quantities display an exponential decay characterized by a typical
temperature dependent length, see Fig. \ref{fig4}. The data have been obtained for a system with $640 \times 320$ spins
after averaging over at least fifty independent runs. Error
bars are smaller than the size of the symbols.
}
\end{figure}
%%%%%%%%%%%%%%%%%%%%%%%%%%%%%%%%%%%%%%%%%%%FIG 3.%%%%%%%%%%%%%%%%%%%%%%%%%%%%%%%%%%%%%%%%%%%%%%%%%%%%%%

%%%%%%%%%%%%%%%%%%%%%%%%%%%%%%%%%%%%%%%%%%%FIG 4.%%%%%%%%%%%%%%%%%%%%%%%%%%%%%%%%%%%%%%%%%%%%%%%%%%%%%%
\begin{figure} [h]
\includegraphics[width=0.80\columnwidth]{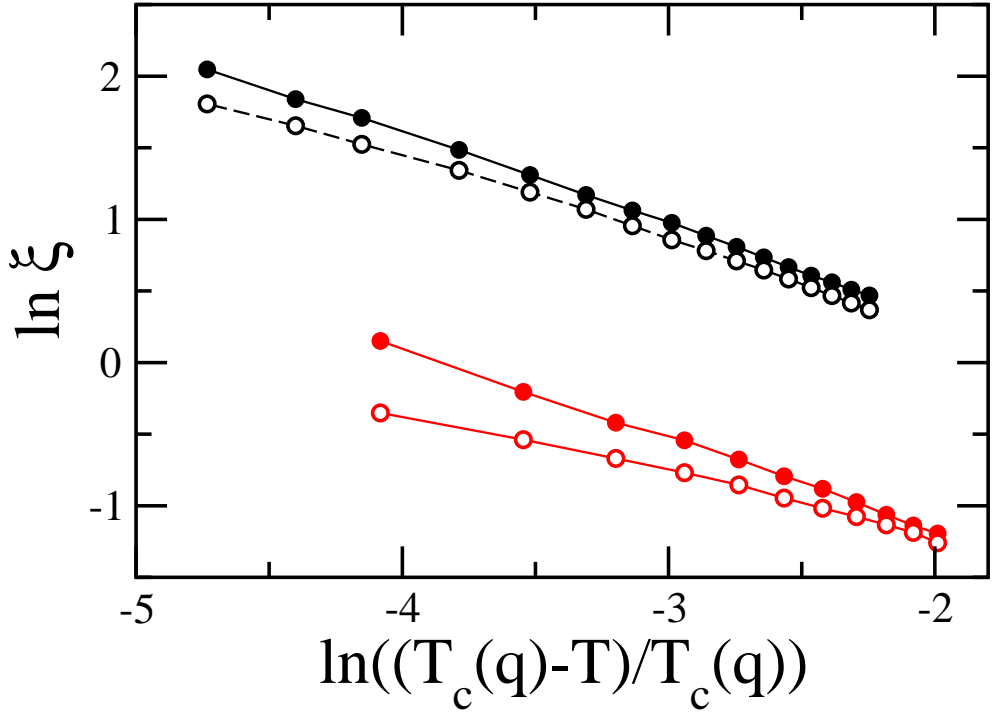}
\caption{\label{fig4} (Color online)
Temperature dependent correlation lengths as obtained from the spatial correlation functions, see
Fig. \ref{fig3} for examples. The black (red) lines are for the $q=9$ ($q=100$) system, with filled (open)
symbols indicating the correlation length parallel (perpendicular) to the surface. On approaching the bulk transition
temperature the correlation lengths display an algebraic behavior. The values 
of the correlation length exponents obtained from the slopes in this figure are collected in Table I.
}
\end{figure}
%%%%%%%%%%%%%%%%%%%%%%%%%%%%%%%%%%%%%%%%%%%FIG 4.%%%%%%%%%%%%%%%%%%%%%%%%%%%%%%%%%%%%%%%%%%%%%%%%%%%%%%

In order to capture the expected anisotropy of the critical surface we study the surface-surface and
surface-bulk correlations, see Eqs. (\ref{Css}) and (\ref{Csb}). As shown in Fig. \ref{fig3} for the
example of $q=9$ states these spatial correlations, after some short-distance regime, rapidly display an
exponential dependence on the distance. The corresponding correlation lengths, see Eqs. (\ref{Css})-(\ref{xiperp}),
increase with temperature, as shown in Fig. \ref{fig4} for the $q=9$ and $q=100$ cases.
This figure also reveals that we are indeed dealing with strongly anisotropic critical systems,
governed by direction dependent correlation length critical exponents, see Table I
for our estimates for these exponents.

Table I provides strong indications that the surface critical exponents at the first-order
phase transition depend on the number of states. Further support comes from the fact that we have the
possibility to do some consistency checks. We first notice from Eq. (\ref{consistency}) that 
the values of $\nu_\parallel$ and $\beta_s$ should differ by one: $\nu_\parallel - \beta_s = 1$.
From our estimates in Table I we obtain the following values for this difference: $1.05 \pm 0.06$ for $q=5$,
$1.07 \pm 0.06$ for $q=9$, $1.01 \pm 0.06$ for $q=16$, and $0.99 \pm 0.04$ for $q=100$. Therefore the 
difference between these two exponents is marginally compatible with 1 for the smaller values of $q$ and
in full agreement with the scaling relation (\ref{consistency}) for the larger values of $q$.

\begin{figure} [h]
\includegraphics[width=0.80\columnwidth]{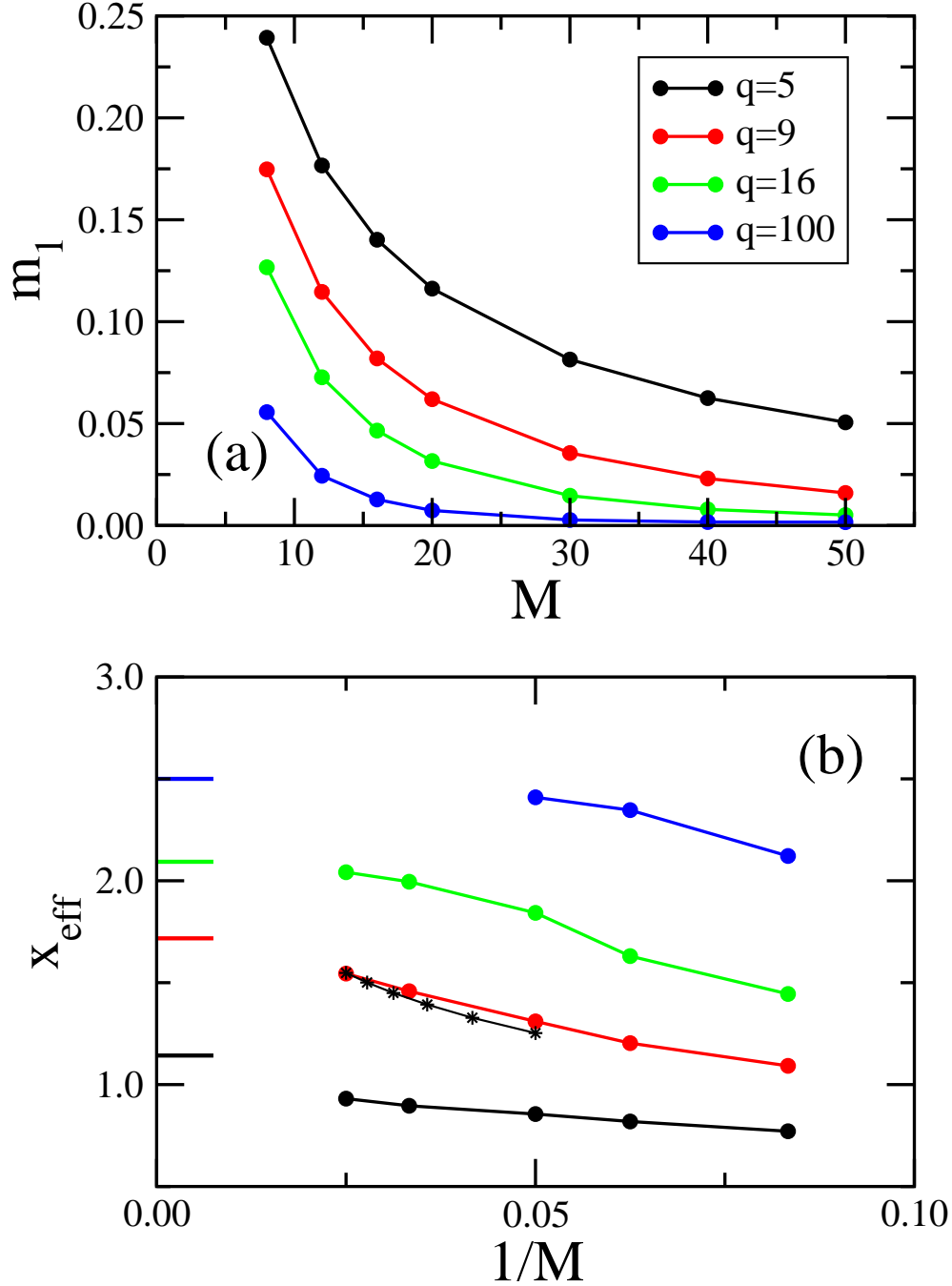}
\caption{\label{fig5} (Color online)
(a) Surface magnetization at the bulk transition temperature for stripes of width $M$ where at one of the surfaces we impose fixed 
boundary conditions. The different curves correspond to different numbers of states $q$. The length of
the system is $L=64000$.
(b) Effective scaling dimension obtained from the data shown in panel (a). 
Only data not affected by finite size effects are
displayed. The bars on the $y$ axis indicate the values obtained from Table I when assuming 
the scaling relation $x = \beta_1/\nu_\perp$. The stars are the data given in Table IV of Ref. \cite{Igl99}
for the $q=9$ case using DMRG.
}
\end{figure}

For a second consistency check we can try to exploit finite-size effects in order to obtain an estimate
of the ratio $x = \beta_1/\nu_\perp$. We thereby follow \cite{Igl99} and consider at the bulk
transition temperature $T_c(q)$ systems with mixed boundary conditions
where at one edge we keep all the spins at a common value, whereas the other edge is a free edge.
For the $q=7$, 8, and 9 cases
Igl\'{o}i and Carlon used the DMRG method in order to calculate the surface magnetization as a function
of the width $M$ of the system and obtained estimates for
the scaling dimension $x = \beta_1/\nu_\perp$ for the $\infty \times M$ system, with $M$ going up to 40. 
In our Monte Carlo simulations we can only simulate finite $L$. Going up to L=64000 sites, we carefully monitor
finite-size effects in that direction. Fig. \ref{fig5}a shows the resulting data for 
the surface magnetization as a function of the stripe
width $M$ for $L=64000$.

%%%%%%%%%%%%%%%%%%%%%%%%%%%%%%%%%%%%%%%%%%%FIG 5.%%%%%%%%%%%%%%%%%%%%%%%%%%%%%%%%%%%%%%%%%%%%%%%%%%%%%%
%%%%%%%%%%%%%%%%%%%%%%%%%%%%%%%%%%%%%%%%%%%FIG 5.%%%%%%%%%%%%%%%%%%%%%%%%%%%%%%%%%%%%%%%%%%%%%%%%%%%%%%

We can obtain an effective exponent $x_{eff}$ from these data through the following logarithmic derivative:
\begin{equation}
x_{eff}(M) = - \frac{d \ln m_1}{d \ln M}~.
\end{equation}
Comparing effective exponents for different values of $L$, only those data not affected by finite-size effects
have been included in Fig. \ref{fig5}b. We also include as stars the data obtained in \cite{Igl99} for the 
$q=9$ case using DMRG, see Table IV in that paper. We note that for $q=100$ we are only able to achieve data representative
for $L =\infty$ for small values of $M$.
This figure shows as bars on the $y$ axis the estimates for the scaling 
dimension $x$ that result when using the values for $\beta_1$ and $\nu_\perp$ given in the Table I. The consistency
of these estimates with the effective exponents lend additional support to a $q$ dependence of the surface
critical exponents at surface induced disordering.

\section{Discussion and conclusion}
Systems undergoing discontinuous bulk transitions may exhibit diverging correlation lengths at surfaces due
to surface induced disordering. While a variety of theoretical studies have verified this general scenario,
only very few attempts have been made to determine directly surface critical exponents through numerical simulations.
The results have been somehow contradictory. Whereas DMRG results for stripes of Potts systems \cite{Igl99} seemed to indicate
a high degree of universality, with surface exponents that do not depend on the number of states $q$,
simulations of face-centered-cubic \cite{Sch96}
and body-centered-cubic \cite{Haa00} alloys found nonuniversal surface exponents.

In this work we have revisited the two-dimensional $q$ states Potts model through extensive numerical simulations.
Focusing on systems with $q=5$, 9, 16, and 100 states, we measured the time-dependent magnetization profiles as well as
surface-surface and surface-bulk correlation functions. The extracted values of surface critical exponents indicate
that the exponents are not universal but depend on the number of states $q$, see Table I. We verified that our data fulfill a
scaling relation connecting the exponent of the surface excess magnetization with the correlation length exponent for
correlations parallel to the surface. In addition, data obtained at the bulk transition temperature
for systems with mixed boundary conditions are found to yield 
effective scaling dimensions that are compatible with the exponents obtained from the temperature dependent quantities.
It is this consistency of our data that allows us to make the case that the values of surface critical exponents 
at the first-order transition of the two-dimensional Potts model depend on the number of states $q$.

Our results, together with those obtained in \cite{Sch96,Haa00} for the three-dimensional alloys, indicate that
the local critical exponents at surface induced disordering are not yet completely understood. New insights could
come from field-theoretical treatments along lines similar to those used to study surface criticality at a
continuous phase transition \cite{Die86}. We hope that our work will lead to future studies in that direction.

\begin{acknowledgments}
This work is supported by the US National
Science Foundation through grant DMR-1205309.

\end{acknowledgments}

\end{document}